\documentstyle[aps,prd,preprint]{revtex}
\def\fun#1#2{\lower3.6pt\vbox{\baselineskip0pt
\lineskip.9pt\ialign{$\mathsurround=0pt#1\hfill##
\hfil$\crcr#2\crcr\sim\crcr}}}

\begin{document}
\vspace{0.5in}
\title{\vskip-2.5truecm{\hfill \baselineskip 14pt 
{\hfill {\small \hfill UT-STPD-6/98 }}}\\
\vskip .1truecm \vspace{1.0cm} \vskip 0.1truecm 
{\bf Degenerate Neutrinos and Supersymmetric Inflation}}
\vspace{1cm}
\author{{G. Lazarides}\thanks{lazaride@eng.auth.gr}}
\vspace{1.0cm}
\address{{\it Physics Division, School of Technology,
\\ 
Aristotle University of Thessaloniki,\\
Thessaloniki GR 540 06, \\
Greece.}}
\maketitle
\vspace{2cm}

\begin{abstract}
\baselineskip 12pt

\par
A moderate extension of the minimal supersymmetric standard 
model which includes a $U(1)_{B-L}$ gauge group ($B$ and 
$L$ being the baryon and lepton number) and a Peccei-Quinn 
symmetry, $U(1)_{PQ}$~, is presented. The hybrid 
inflationary scenario is automatic and `natural' in this 
model. The $\mu$ problem of the minimal supersymmetric 
standard model is solved by coupling the electroweak 
higgses to fields which break $U(1)_{PQ}$~. Baryon number 
conservation and, thus, proton stability are automatic 
consequences of a R-symmetry.  Neutrinos are assumed 
to acquire degenerate masses $\approx 1.5~{\rm{eV}}$ 
by coupling to $SU(2)_L$ triplet superfields, thereby 
providing the hot dark matter of the universe. The inflaton 
system decays into these triplets which, via their subsequent 
decay, produce a primordial lepton asymmetry later converted 
into the observed baryon asymmetry of the universe. The 
gravitino and baryogenesis constraints can be satisfied with 
`natural' values ($\sim 10^{-3}$) of the relevant coupling 
constants.       
\end{abstract}

\thispagestyle{empty} 
\newpage 
\pagestyle{plain} 
\setcounter{page}{1}
\baselineskip 20pt

\par
It is well-known that the minimal supersymmetric standard 
model (MSSM), despite its compelling properties, leaves 
a number of fundamental questions unanswered. This  
clearly indicates that MSSM must be part of a larger 
scheme. One shortcoming of MSSM is that inflation cannot 
be implemented in its context. Also, there is no
understanding of how the supersymmetric $\mu$ term, with 
$\mu \sim 10^{2}-10^{3}~{\rm{GeV}}$, arises. It has 
become increasingly clear that a combination of both cold 
and hot dark matter provides \cite{structure} a good fit 
to the data on large scale structure formation in the 
universe. Although the lightest supersymmetric particle 
of MSSM is a promising candidate for cold dark matter, 
hot dark matter cannot be accommodated with purely MSSM 
fields. Finally, the observed baryon asymmetry of the 
universe cannot be generated easily in MSSM through the 
electroweak sphaleron processes.

\par
A moderate extension of MSSM based on the gauge group 
$G_{S}\times U(1)_{B-L}$ ($G_S$ being the standard 
model gauge group, and $B$ and $L$ the baryon and 
lepton number) provides \cite{lyth,dss} a suitable 
supersymmetric framework for inflation. Indeed, the 
hybrid inflationary scenario \cite{hybrid} is `naturally' 
realized in this context. Inflation is associated with 
the breaking of $U(1)_{B-L}$ at a superheavy scale. The 
$\mu$ problem of MSSM could be solved \cite{dls} by 
coupling the inflaton to the electroweak higgses. However, 
in this case, the inflaton predominantly decays into higgs 
superfields, after the end of inflation,  and the gravitino 
constraint \cite{gravitino} on the reheat temperature
restricts \cite{atmospheric} the relevant dimensionless 
coupling constants to be `unnaturally' small 
($\sim 10^{-5}$). We will, thus, choose here an 
alternative solution \cite{kn,rsym} to the $\mu$ problem. 
This relies on the coupling of the electroweak 
higgses to superfields causing the breaking of a Peccei-Quinn 
\cite{pq} symmetry rather than to the inflaton.

\par
The hot dark matter of the universe, needed for explaining 
\cite{structure} its large scale structure, may be 
provided by light neutrinos. This possibility can be made 
compatible with the atmospheric \cite{superk} and solar 
neutrino oscillations, within a three neutrino scheme, only 
by assuming almost degenerate neutrino masses. These masses 
can be generated by including \cite{rsym,triplet} $SU(2)_L$ 
triplet pairs of superfields with intermediate scale masses. 
The inflaton decays into these triplet superfields (rather 
than into higgses) and much bigger dimensionless coupling 
constants are allowed. The subsequent decay of the triplet 
superfields produces \cite{sarkar} a primordial lepton 
asymmetry \cite{leptogenesis} which is later converted 
into the observed baryon asymmetry of the universe by 
electroweak sphaleron effects.
  
\par
Let us now describe, in some detail, the moderate extension 
of MSSM based on the gauge group $G= G_S\times 
U(1)_{B-L}$~. The spontaneous breaking of $U(1)_{B-L}$ at 
a mass scale $M\sim 10^{16}~{\rm{GeV}}$ is achieved 
through the renormalizable superpotential 
\begin{equation}
W=\kappa S(\phi\bar{\phi}-M^2)~,  
\label{W}
\end{equation}
where $\phi,~\bar{\phi}$ is a conjugate pair of standard 
model singlet left handed superfields with $B-L$ charges 
equal to 1, -1 respectively, and $S$ is a gauge singlet left 
handed superfield. The coupling constant $\kappa$ and the 
mass parameter $M$ can be made real and positive by 
suitable redefinitions of the phases of the superfields. 
The supersymmetric minima of the scalar potential 
lie on the D-flat direction $\phi=\bar{\phi}^*$ at 
$\langle S\rangle = 0~,~ |\langle \phi \rangle| 
=|\langle \bar{\phi} \rangle| = M$.

\par
It has been well documented 
\cite{lyth,dss,dls,atmospheric,lss} that hybrid 
inflation \cite{hybrid} is automatically 
and `naturally' realized in this supersymmetric scheme. 
The scalar potential possesses a built-in inflationary 
trajectory at $\phi=\bar{\phi}=0$~, 
$|S|>M$ with a constant tree level potential energy 
density $\kappa^{2}M^{4}$ which is responsible for 
the exponential expansion of the universe. 
Moreover, due to supersymmetry breaking by this constant 
energy density, there are important radiative corrections 
\cite{dss} which provide a slope along the inflationary 
trajectory necessary for driving the inflaton towards the 
supersymmetric vacua. At one loop, the cosmic microwave 
quadrupole anisotropy is given by \cite{dss,lss}  
\begin{equation}
\left(\frac{\delta T}{T}\right)_{Q}\approx 8\pi
\left(\frac{N_{Q}}{45}\right)^{1/2}\frac{x_{Q}}
{y_{Q}}\left(\frac{M}{M_{P}}\right)^{2}.
\label{quadrupole}
\end{equation} 
Here $M_{P}=1.22\times 10^{19}~{\rm{GeV}}$ is the 
Planck scale and $N_Q \approx 50-60$ denotes the number 
of e-foldings experienced by the universe between the 
time the quadrupole scale exited the horizon and the end 
of inflation. Also, 
\begin{eqnarray*}
y_Q^2=\int_{1}^{x_Q^{2}}\frac{dz}{z((z-1)
\ln (1-z^{-1})+(z+1)\ln (1+z^{-1}))}
\end{eqnarray*}
\begin{equation}
=x_Q^2\left(1-\frac{7}{6x_Q^2}+\cdots\right)~,
~y_Q \geq 0~,
\label{yQ}
\end{equation}
with $x_{Q}=|S_{Q}|/M$ ($x_{Q}\geq 1$), $S_{Q}$ 
being the value of the scalar field $S$ when the scale 
which evolved to the present horizon size crossed outside 
the de Sitter (inflationary) horizon. Note that 
Eq.(\ref{quadrupole}) holds, to a good approximation, 
provided $x_{Q}$ is not `unnaturally' close to 1. The 
superpotential parameter $\kappa$ can be evaluated 
\cite{dss,lss} from
\begin{equation} 
\kappa \approx \frac{8\pi ^{3/2}}
{\sqrt{N_{Q}}}~y_{Q}~\frac{M}{M_{P}}~\cdot
\label{kappa}
\end{equation}

\par
One interesting possibility for generating the $\mu$ term 
of MSSM has been proposed in Ref.\cite{dls}. It relies on 
the extension of the above scheme by adding to it the 
superpotential coupling $\lambda S H^{(1)}H^{(2)}$ 
($\lambda>\kappa$), where $H^{(1)}, H^{(2)}$ are the 
chiral higgs superfields which couple to the up and down 
type quarks respectively (and carry zero $B-L$ charge). 
It has been shown \cite{dls} that, after gravity-mediated 
supersymmetry breaking, $S$ develops a vacuum expectation 
value (vev) $\langle S\rangle\approx -m_{3/2}/\kappa$~, 
where $m_{3/2}\sim (0.1-1)\ {\rm TeV}$ is the gravitino 
mass. This generates a $\mu$ term with $\mu=\lambda 
\langle S\rangle\approx -(\lambda/\kappa)m_{3/2}$~. 
The cosmic microwave quadrupole anisotropy, evaluated (see 
Ref.\cite{atmospheric}) in the limit $x_Q\rightarrow 1$ 
which is the relevant one here, is given by 
\begin{equation}
\left(\frac{\delta T}{T}\right)_{Q}\approx 
\frac{32\pi^{5/2}}{3\sqrt{5}}
\left(\frac{M}{M_P}\right)^{3}
\frac{1}{\kappa(\epsilon\frac{\lambda^2}
{\kappa^2}+\ln{2})}~, 
\label{anisotropy}
\end{equation}
where $\epsilon \approx 2\ln{2}$ for 
$\lambda\approx \kappa$, and  $\epsilon \approx 1$ 
for $\lambda \gg \kappa$. (Notice that here we had to 
replace the contribution of the conjugate pair of 
$SU(2)_R$ doublet superfields of Ref.\cite{atmospheric}
by the contribution of the standard model singlets $\phi$, 
$\bar{\phi}$, which is smaller by a factor 2.) After the 
end of inflation, the inflaton (oscillating system), 
which consists of the two complex scalar fields $S$ and 
$\theta=(\delta \phi + \delta\bar{\phi})/\sqrt{2}$ 
($\delta \phi=\phi-M$, $\delta \bar{\phi}=
\bar{\phi}-M$) with mass $m_{infl}=\sqrt{2}\kappa M$, 
predominantly decays, in this case, into ordinary 
higgsinos and higgses with a decay width $\Gamma=
(1/8\pi)\lambda^{2}m_{infl}$. This can be easily 
deduced from the coupling $\lambda S H^{(1)}H^{(2)}$ 
and the superpotential in Eq.(\ref{W}). The reheat 
temperature is given \cite{lss} by
\begin{equation}
T_r\ \approx \frac{1}{7} 
\left( \Gamma M_P\right)^{1/2}~, 
\label{reheat}
\end{equation}
for MSSM spectrum. Using Eqs.(\ref{anisotropy}) and 
(\ref{reheat}) with $(\delta T/T)_{Q}\approx 
6.6\times 10^{-6}$ from the Cosmic
Background Explorer (COBE) \cite{cobe}, the gravitino 
constraint \cite{gravitino} 
($T_r\stackrel{_{<}}{_{\sim }}10^9$ GeV) becomes
\begin{equation}
\lambda\kappa^{2/3}\left(\epsilon\frac{\lambda^2}
{\kappa^2}+\ln{2}\right)^{1/6}
\stackrel{_{<}}{_{\sim }}3.7\times 10^{-8}~,
\label{bound}
\end{equation}
which, for $\lambda=\kappa$ say, gives $\kappa
\stackrel{_{<}}{_{\sim }}3.2\times 10^{-5}$. In 
the specific model of Ref.\cite{atmospheric}, $\lambda 
\approx 3.95\kappa$ and the bound is even stronger, 
i.e., $\kappa\stackrel{_{<}}{_{\sim }}1.2
\times 10^{-5}$. Moreover, in this model which employs 
hierarchical neutrino masses from the seesaw mechanism,
the requirement of maximal $\nu_{\mu}$-$\nu_{\tau}$ 
mixing, deduced from the recent results of the 
SuperKamiokande experiment \cite{superk}, further reduces 
the coupling constant $\kappa$ to become of order $10^{-6}$.
We conclude that, within the context of the supersymmetric 
hybrid inflationary model, the solution of the $\mu$ problem 
of MSSM via the coupling of the higgs superfields to the 
inflaton system is not totally satisfying. The reason is that 
this solution together with the gravitino constraint 
require the presence of `unnaturally' small coupling 
constants.

\par
We will follow here an alternative solution \cite{kn} of the 
$\mu$ problem of MSSM constructed by coupling the electroweak 
higgses to superfields causing the breaking of a Peccei-Quinn 
\cite{pq} symmetry ($U(1)_{PQ}$) which solves the strong 
CP problem. D- and F-flat directions in field space, 
appearing in the supergravity extension of MSSM, can 
generate an intermediate scale $M_I$ which, in the simplest 
case, is given by $M_I \sim (m_{3/2}m_P)^{1/2} \sim 
10^{11}~{\rm{GeV}}~$, where $m_P=M_P/\sqrt{8\pi} \approx 
2.44\times 10^{18}~{\rm{GeV}}$ is the `reduced' Planck mass. 
This scale can be identified with the symmetry breaking scale 
$f_{a}$ of $U(1)_{PQ}$~. A $\mu$ term with $\mu \sim 
m_{3/2} \sim f^2_{a}/m_P$ can then be easily generated 
\cite{kn} via an appropriate non-renormalizable coupling of 
the electroweak higgses to a field which breaks $U(1)_{PQ}$~.

\par
We will assume almost degenerate (rather than 
hierarchical) light neutrino masses. Under this assumption, 
neutrinos can provide the hot dark matter of the universe 
which is needed for explaining \cite{structure} its large 
scale structure. In the hierarchical case, atmospheric and 
solar neutrino oscillations imply that, within a three 
neutrino scheme, the neutrinos have too small masses to be 
of any cosmological significance. Non-zero neutrino masses 
can be generated by introducing into the scheme standard 
model singlet right handed neutrinos and/or by including 
\cite{rsym,triplet} $SU(2)_L$ triplet pairs of 
superfields. The former possibility, which is based on 
the well-known seesaw mechanism, cannot naturally lead 
to degenerate neutrino masses. We, thus, adopt here the 
latter option. The $SU(2)_L$ triplets acquire intermediate 
masses via non-renormalizable superpotential couplings to 
the inflaton. These same couplings, which are automatically 
suppressed by a factor $M/m_P$, cause the decay of the 
inflaton predominantly into these triplet superfields 
(rather than into higgses). This allows much 
bigger dimensionless coupling constants, thereby solving 
the `naturality' problem of the previous scheme. The 
subsequent decay of the triplet superfields produces 
\cite{sarkar} a primordial lepton asymmetry 
\cite{leptogenesis} which is later converted into the 
observed baryon asymmetry of the universe by electroweak 
sphaleron effects. For this to work, we need at least two 
pairs of triplets.

\par
We now proceed to the description of the full model. 
We supplement the spectrum of the moderate extension
of MSSM in Eq.(\ref{W}), which incorporates hybrid 
inflation, with a pair of gauge singlet left handed 
superfields $N,~\bar{N}$. Their vevs will break 
$U(1)_{PQ}$ at an intermediate scale. We also add two 
pairs of $SU(2)_L$ triplets $T_a,~\bar{T}_a$ 
($a$=1,2) with hypercharges 1, -1 and $B-L$ charges 
2, 0 respectively. They will be responsible for neutrino 
masses and the generation of the primordial lepton 
asymmetry. The superpotential $W$ contains, in addition 
to the terms in Eq.(\ref{W}), the following couplings:
\begin{equation}
H^{(1)}QU^c,~H^{(2)}QD^c,~H^{(2)}LE^c,
~N^{2}H^{(1)}H^{(2)},~N^{2} \bar{N}^2,
~TLL,~\bar{T} H^{(1)} H^{(1)},
~\bar{\phi}\bar{\phi}T\bar{T}~.
\label{couplings}
\end{equation}
Here $Q_i$ denote the $SU(2)_L$ doublet quark 
superfields, $U^c_i$ and $D^c_i$ are the $SU(2)_L$ 
singlet quark superfields, while $L_i~(E^c_i)$ stand 
for the $SU(2)_L$ doublet (singlet) lepton superfields 
($i$=1,2,3 is the family index). The $B-L$ charges 
of these fields are defined in the obvious way. The 
quartic terms in Eq.(\ref{couplings}) carry a factor 
$m^{-1}_P$ which has been left out. Also, the 
dimensionless coupling constants as well as the 
family and triplet indices are suppressed. 

\par
The continuous global symmetries of this superpotential 
are $U(1)_B$ (and, consequently, $U(1)_L$) with the 
extra chiral superfields $S$, $\phi$, $\bar{\phi}$, 
$N$, $\bar{N}$, $T$, $\bar{T}$ carrying zero baryon 
number, an anomalous Peccei-Quinn symmetry $U(1)_{PQ}$, 
and a non-anomalous R-symmetry $U(1)_{R}~$. The $PQ$ and 
$R$ charges of the various superfields are as follows:
\begin{equation}
\begin{array}{rcl}
PQ:~H^{(1)}(1),~H^{(2)}(1),
~Q(-1),~U^c(0),~D^c(0),~L(-1),~E^c(0),~~~~
\\
~S(0),~\phi(0),~\bar{\phi}(0),
~N(-1),~\bar{N}(1),~T(2),~\bar{T}(-2)~,
~~~~~~~~~~~~~ 
\\
& & \\
R:~H^{(1)}(0),~H^{(2)}(0),~Q(1/2),~U^c(1/2),
~D^c(1/2),~L(1/2),~E^c(1/2),
\\
~S(1),~\phi(0),~\bar{\phi}(0),
~N(1/2),~\bar{N}(0),~T(0),~\bar{T}(1)~,
~~~~~~~~~~~~~~
\end{array}
\label{charges}
\end{equation}
with $W$ carrying one unit of $R$ charge.

\par
It is important to note that $U(1)_B$ (and, consequently, 
$U(1)_L$) is automatically implied by $U(1)_{R}$ even if 
all possible non-renormalizable terms are included in the 
superpotential. Indeed, by extending the $U(1)_{R}$ symmetry 
to higher order terms, one can show that $U(1)_B$ follows as 
a consequence. To see this, observe that the $R$ charges of 
the only baryon number violating combinations of fields 
$3 \cdot 3 \cdot 3$ or 
$\bar{3} \cdot \bar{3} \cdot \bar{3}$ (3, $\bar{3}$ 
denote color triplet and antitriplet fields) 
exceed unity and cannot be compensated since there are no 
negative $R$ charges available in the model. In particular, 
the troublesome dimension five operators $QQQL$ and 
$U^cU^cD^cE^c$ are eliminated and proton is stable.

\par
Lepton number is spontaneously broken by the vevs of 
$\phi$, $\bar{\phi}$ and, consequently, some lepton number 
violating effective operators will emerge at lower energies 
(below $M$). In particular, the last term in 
Eq.(\ref{couplings}) will generate the desired intermediate 
scale masses for the $SU(2)_L$ triplet superfields. However, 
undesired mixing of the higgs $H^{(2)}$ with $L$ 's will 
also emerge from the allowed superpotential couplings 
$N\bar{N}LH^{(1)}\phi$ after the breaking of $U(1)_{PQ}$ 
by the vevs of $N$, $\bar{N}$. To avoid this complication,
we impose an extra discrete $Z_2$ symmetry, which we will 
call `lepton parity'. Under this symmetry, $L$, $E^c$ change 
sign, while all other superfields remain unaltered. In the 
present model, this symmetry is equivalent with 
`matter parity' (under which $L$, $E^c$, $Q$, $U^c$, $D^c$ 
change sign), since `baryon parity' (under which $Q$, $U^c$, 
$D^c$ change sign) is also present being a subgroup of 
$U(1)_B$~. One can show that the only superpotential terms, 
which are permitted by the global symmetries $U(1)_{R}$~, 
$U(1)_{PQ}$ and `lepton parity', are the ones of 
Eqs.(\ref{W}) and (\ref{couplings}) modulo arbitrary 
multiplications by non-negative powers of the combination 
$\phi\bar{\phi}$.

\par
The scalar potential which is generated by the 
superpotential term $N^{2} \bar{N}^2$ in 
Eq.(\ref{couplings}) after gravity-mediated 
supersymmetry breaking has been studied in 
Ref.\cite{rsym}. It has been shown that, for a 
suitable choice of parameters, a minimum at 
\begin{equation}
|\langle N \rangle|~=~|\langle\bar{N}\rangle|
~\sim (m_{3/2}m_{P})^{1/2}
\label{vev}
\end{equation}
is preferred over the trivial one at 
$\langle N\rangle=\langle\bar{N}\rangle=0$.
The vevs $\langle N \rangle,~\langle\bar{N}\rangle$  
together break $U(1)_{PQ} \times U(1)_{R}$ completely. 
Substitution of these vevs in the superpotential coupling 
$N^{2}H^{(1)}H^{(2)}$ in Eq.(\ref{couplings}) then 
generates a $\mu$ parameter for MSSM of order $m_{3/2}$ 
as desired. Note that $U(1)_{L}$ is broken completely 
together with the gauge $U(1)_{B-L}$ by the superheavy 
vevs of $\phi$, $\bar{\phi}$. Thus, only $U(1)_{B}$ 
and `matter parity' remain exact.

\par
As already explained, after $B-L$ (and lepton number) 
breaking at the superheavy scale $M$, the last term in 
Eq.(\ref{couplings}) generates intermediate scale 
masses for the $SU(2)_L$ triplet superfields $T_a$, 
$\bar{T}_a$ ($a$=1,2). The dimensionless coupling 
constant matrix of this term can be made diagonal with 
positive entries $\gamma_a$ ($a$=1,2) by a suitable 
rotation on the triplets. The triplet mass eigenvalues 
are then $M_a=\gamma_a M^{2}/m_{P}$ (with $\langle 
\phi \rangle$, $\langle \bar{\phi} \rangle$ taken 
positive by an appropriate $B-L$ transformation). It is 
readily checked that the scalar components of $T_a$ 's 
acquire non-zero vevs $\sim M^2_W/M~(\ll M_W)$, with 
the electroweak breaking playing an essential role in 
the generation of these vevs. This is due to the fact that 
the last two terms in Eq.(\ref{couplings}), after 
electroweak breaking, give rise to terms linear with 
respect to $T_a$ 's in the scalar potential of the 
theory. The vev of $T_a$ is then given by $\langle T_a
\rangle=\beta_{a}\langle H^{(1)}\rangle^{2}/M_{a}$~,
with $\beta_{a}$ being the dimensionless coupling 
constant of the term $\bar{T}_aH^{(1)} H^{(1)}$. 
These vevs violate lepton number and generate a non-zero 
mass matrix for neutrinos, $m_{\nu}=\sum_{a=1,2}
\alpha_{aij}\beta_{a}\langle H^{(1)} \rangle^{2}/
M_a~$, via the term $\alpha_{aij}T_{a}L_{i}L_{j}$. 
Note that $U(1)_B$ and `matter parity' still survive 
as exact symmetries. The neutrino mass matrix  
can be diagonalized by a suitable `Kobayashi-Maskawa' 
rotation in its standard form (involving three angles 
and a CP violating phase) and the complex eigenvalues 
can be written as
\begin{equation} 
m_i=\sum_{a=1,2}\alpha_{ai}\beta_{a}
\frac{\langle H^{(1)}\rangle^{2}}{M_a}~,
\label{mass}
\end{equation}
where $\alpha_{ai}$ are the (complex) eigenvalues 
of the complex symmetric matrices $\hat{\alpha}_{a}
=(\alpha_{aij})$. Note that the $m_i$ 's, being in 
general complex, carry two extra CP violating phases 
(an overall phase factor is irrelevant) which appear 
in some processes like double-beta decay.

\par
For definiteness, we will adopt the model of neutrino 
masses and mixing discussed in Ref.\cite{gg}. This scheme 
has almost degenerate neutrino masses and employs the 
bimaximal neutrino mixing \cite{bimaximal} which is 
consistent with the vacuum oscillation explanation 
\cite{vacuum} of the solar neutrino puzzle. Moreover, 
all three neutrino masses are real, but the CP parity of 
one of them (say the second one) is opposite to the CP 
parities of the other two. This is important for satisfying 
the experimental constraints \cite{beta} from neutrinoless 
double beta decay. Although favored by data, this scheme 
has not been derived so far from a simple set of 
symmetries. (One interesting attempt with four neutrinos 
appeared in Ref.\cite{mohapatra}).

\par
We will not undertake here the ambitious task of 
implementing the above scheme of neutrino masses and 
mixing in the context of our model. We will restrict 
ourselves to observing that the required neutrino mass 
parameters can be obtained in our model provided the 
coupling constants $\alpha_{ai}$ ($a$=1,2; $i$=1,2,3) 
satisfy the relations $\alpha_{a1}=-\alpha_{a2}=
\alpha_{a3}\equiv\alpha_{a}$ to a very good 
approximation. The precise values of mixing angles and 
square-mass differences turn out to be irrelevant for 
our purposes.

\par
We now turn to the discussion of the decay of the inflaton, 
which consists of the two complex scalar fields $S$ and 
$\theta$. The scalar $\theta$ ($S$) can decay into a 
pair of fermionic (bosonic) $T_a$, 
$\bar{T}_a$ 's, as one easily deduces from the last 
coupling in Eq.(\ref{couplings}) and the coupling 
$\kappa S\phi\bar{\phi}$. The decay width is the same 
for both scalars and equals
\begin{equation} 
\Gamma=\frac{3}{8\pi}~\gamma_{a}^{2}
\left(\frac{M}{m_P}\right)^{2}m_{infl}~.
\label{width}
\end{equation}
Of course, decay of the inflaton into   
$T_a$, $\bar{T}_a$ is possible provided the corresponding
triplet mass $M_a \leq m_{infl}/2$. The gravitino 
constraint \cite{gravitino} on the reheat temperature, 
$T_r$, then implies strong bounds on the $M_a$ 's which 
satisfy this inequality. Consequently, the corresponding 
dimensionless coupling constants, $\gamma_{a}$~, are 
restricted to be quite small.

\par
To minimize the number of small couplings, we then take 
$M_2 < m_{infl}/2\leq M_1$ so that the inflaton 
decays into only one (the lightest) triplet pair with 
mass $M_2$. Moreover, we take $\gamma_{1}=1$, which 
gives $M_{1}=M^{2}/m_{P}$ and allows us to maximize the 
parameter $\kappa$ (see below). Using 
Eq.(\ref{kappa}), the requirement $m_{infl}/2\leq M_1$
becomes $y_{Q}\leq \sqrt{N_{Q}}/2\pi \approx 1.19$,
for $N_{Q}=56$, and Eq.(\ref{yQ}) gives 
$x_{Q}\leq 1.59$. To maximize $\kappa$ (and $M$), we 
choose $x_{Q}=1.59$. Eqs.(\ref{quadrupole}), 
(\ref{kappa}) with $(\delta T/T)_{Q}\approx 6.6
\times 10^{-6}$ from COBE \cite{cobe}  
then give $M \approx 5.12\times 10^{15}~ 
{\rm{GeV}}$, $\kappa \approx 2.97\times 10^{-3}$. Also, 
the inflaton mass is $m_{infl} \approx 2.15\times 10^{13}
~{\rm{GeV}}$, the triplet masses are  $M_{1} \approx 
1.07 \times 10^{13}~{\rm{GeV}}$, $M_{2} \approx 1.07
\gamma_{2}\times 10^{13}~{\rm{GeV}}$, and the reheat 
temperature is $T_{r} \approx 1.68\gamma_{2}\times 
10^{12}~{\rm{GeV}}$. The gravitino constraint 
\cite{gravitino}
($T_r\stackrel{_{<}}{_{\sim }}10^9$ GeV) then implies 
$\gamma_{2}\stackrel{_{<}}{_{\sim }} 5.96 
\times 10^{-4}$ which is somewhat small.
We do not consider this bound on the coefficient of a 
non-renormalizable superpotential coupling unacceptable. 
However, larger $\gamma_{2}$ 's 
can be accommodated by allowing bigger $T_{r}$ 's which 
is possible provided the branching ratio of gravitinos 
to photons is small enough 
(see Ref.\cite{photon}). Larger $\gamma_{2}$ 's 
can also be obtained without relaxing the gravitino 
constraint, but at the expense of having smaller 
$\kappa$ 's. For example, taking $x_{Q}=1.2$, we 
obtain $y_{Q}=0.61$,  $M \approx 4.22\times 10^{15}~ 
{\rm{GeV}}$, $\kappa \approx 1.26\times 10^{-3}$, 
$m_{infl} \approx 7.53\times 10^{12}~{\rm{GeV}}$,  
$M_{1} \approx 7.31\times 10^{12}~{\rm{GeV}}$, 
$M_{2} \approx 7.31\gamma_{2}\times 10^{12}~
{\rm{GeV}}$, and $T_{r} \approx 8.19\gamma_{2}
\times 10^{11}~{\rm{GeV}}$. The gravitino constraint 
then implies $\gamma_{2}\stackrel{_{<}}{_{\sim }}
1.22\times 10^{-3}$.

\par
Baryon number is violated only  by `tiny' non-perturbative 
$SU(2)_L$ instanton effects in the present scheme. So the 
only way to produce the observed baryon asymmetry of the 
universe is to first generate a primordial lepton asymmetry 
\cite{leptogenesis} which is then partially converted into 
the baryon asymmetry by the non-perturbative sphaleron 
effects of the electroweak sector. The primordial lepton 
asymmetry is produced via the decay of the 
superfields $T_{2}$, $\bar{T}_{2}$ which emerge as decay 
products of the inflaton. This mechanism for leptogenesis 
has been discussed in Refs.\cite{rsym,sarkar}. The $SU(2)_L$ 
triplet superfields decay either to a pair of $L_i$ 's or 
to a pair of $H^{(1)}$ 's. In the absence of right handed 
neutrinos, the one-loop diagrams which interfere with the 
tree level ones are \cite{sarkar} of the self-energy type 
\cite{covi} with a s-channel exchange of $T_{1}$, 
$\bar{T}_{1}$. The resulting lepton asymmetry is 
\cite{sarkar}
\begin{equation}
\frac {n_{L}}{s} \approx -\frac{3}{8 \pi}~
\frac {T_{r}}{m_{infl}}~
\frac{M_{1}M_{2}}{M_{1}^{2}-M_{2}^{2}}~
\frac{{\rm{Im}}(\beta_{1}^{*}\beta_{2}
{\rm{Tr}}(\hat{\alpha}_{1}^{\dagger}
\hat{\alpha}_{2}))}
{\rm{Tr}(\hat{\alpha}_{2}^{\dagger}
\hat{\alpha}_{2})
+\beta_{2}^{*}\beta_{2}}~,
\label{asymmetry}
\end{equation}
where the $3\times 3$ complex symmetric matrix 
$\hat{\alpha}_{a}$, after diagonalization, 
becomes equal to
${\rm{diag}}(\alpha_{a},-\alpha_{a}, \alpha_{a})$. 
Note that the above formula holds provided \cite{pilaftsis} 
the decay width of $T_{1}$, $\bar{T}_{1}$ is much 
smaller than $(M_{1}^{2}-M_{2}^{2})/M_{2}$, which 
is well satisfied here since $M_{2}\ll M_{1}$. 
For MSSM spectrum, the observed baryon 
asymmetry $n_{B}/s$ is related \cite{ibanez} to 
$n_{L}/s$ by $n_{B}/s=-(28/79)(n_{L}/s)$.
It is important to ensure that the primordial lepton 
asymmetry is not erased by lepton number violating 
$2\rightarrow 2$ scatterings at all temperatures 
between $T_r$ and 100 GeV. This requirement gives 
\cite{ibanez} $m_{\nu_{\tau}}\stackrel{_<}{_\sim} 
10~{\rm{eV}}$ which is readily satisfied.
Using Eqs.(\ref{reheat}), (\ref{width}) and the 
fact that $M_{2}\ll M_{1}$, Eq.(\ref{asymmetry}) 
can be simplified as
\begin{equation}
\left|\frac {n_{L}}{s}\right| \approx 
\frac{9\sqrt{3}}{56 \pi}~
\frac{M}{\sqrt{m_{infl}M_{P}}}~
\gamma_{2}^{2}~
\frac{|{\rm{Im}}(\alpha_{1}^{*}\beta_{1}^{*}
\alpha_{2}\beta_{2})|}
{3|\alpha_{2}|^{2}+|\beta_{2}|^{2}}~\cdot
\label{leptoasym}
\end{equation}

\par
The parameters $\alpha_{a}$, $\beta_{a}$, $\gamma_{a}$
($a$=1,2) are constrained by the requirement that the hot 
dark matter of the universe consists of neutrinos. We 
take the `relative' density of hot dark matter 
$\Omega_{HDM} \approx 0.2$, which is favored
by the structure formation in cold plus hot dark matter 
models \cite{structure}, and $h \approx 0.5$, where $h$ 
is the present value of the Hubble parameter in units of 
$100~\rm{km}~\rm{sec}^{-1}~\rm{Mpc}^{-1}$. The 
common mass of the three light neutrinos is then about 
$1.5~{\rm{eV}}$ and Eq.(\ref{mass}) gives the
constraint
\begin{equation} 
\left|\sum_{a=1,2}\frac{\alpha_{a}\beta_{a}}
{\gamma_{a}}\right| \approx\left(
\frac{M}{7.02\times 10^{15}~{\rm{GeV}}}\right)^{2}
\equiv \xi~,
\label{constraint}
\end{equation}
where $|\langle H^{(1)}\rangle|$ was taken equal to
$174~{\rm{GeV}}$. To maximize, under this constraint, 
the numerator of the last fraction in Eq.(\ref{leptoasym}), 
observe that Eq.(\ref{constraint}) implies
\begin{equation}
|{\rm{Im}}(\delta_{1}^{*}\delta_{2})|^{2}=
\frac{1}{4}
\left(\xi^{2}-(|\delta_{1}|-|\delta_{2}|)^{2}\right) 
\left((|\delta_{1}|+|\delta_{2}|)^{2}-\xi^{2}\right)~,
\label{imaginary}
\end{equation}
where $\delta_{a}=\alpha_{a}\beta_{a}/\gamma_{a}$ 
($a$=1,2). For $|\delta_{1}|+|\delta_{2}|$ fixed,
this expression takes its maximal value $\xi^{2}
\left(\delta^{2}-(\xi/2)^{2}\right)$ at 
$|\delta_{1}|=|\delta_{2}|\equiv \delta$. 
Moreover, for fixed $\delta$, the denominator of the 
last fraction in Eq.(\ref{leptoasym}) is minimized 
at $\sqrt{3}|\alpha_{2}|=|\beta_{2}|$ with minimum 
value $2\sqrt{3}\gamma_{2}\delta$. Putting all these 
together, we obtain
\begin{equation}
\left|\frac {n_{L}}{s}\right| 
\stackrel{_{<}}{_{\sim }}\frac{9}{112 \pi}~
\frac{M}{\sqrt{m_{infl}M_{P}}}~
\gamma_{1}\gamma_{2}^{2}~\xi\left(1-
\frac{\xi^{2}}{4\delta^{2}}\right)^{1/2}
\cdot
\label{maximum}
\end{equation}
To further maximize $n_{L}/s$, we take $\alpha_{1}=
\beta_{1}=1$ (remember $\gamma_{1}=1$ too). This choice 
maximizes $\delta$ which becomes equal to 1. For $x_{Q}=
1.59$, $\xi \approx 0.53$ and the maximal lepton asymmetry 
becomes $\approx 4.14~\gamma_{2}^{2}\times 10^{-3}$.
The low deuterium abundance  constraint \cite{deuterium}, 
$\Omega _{B}h^{2}\approx 0.025$, can then be satisfied 
provided $\gamma_{2}\stackrel{_{>}}{_{\sim }} 
2.57\times 10^{-4}$. So, for $\gamma_{2}$ in the range 
$(2.57-5.96)\times 10^{-4}$, both gravitino and 
baryogenesis restrictions can be met. For $x_{Q}=1.2$, 
$\xi \approx 0.36$ and the maximal lepton asymmetry becomes 
$\approx 4.02~\gamma_{2}^{2}\times 10^{-3}$. The allowed 
range of $\gamma_{2}$ is now $2.61\times 10^{-4}-1.22
\times 10^{-3}$. We see that the required values 
($\sim 10^{-3}$) of the relevant coupling constants are 
`natural'.       

\par
In conclusion, we have presented a moderate extension of 
MSSM by including a $U(1)_{B-L}$ gauge group and a 
Peccei-Quinn symmetry ($U(1)_{PQ}$) which solves the 
strong CP problem. The hybrid inflationary scenario is 
automatically and `naturally' realized. The $\mu$ problem 
of MSSM is solved by coupling the electroweak higgses to 
fields which break $U(1)_{PQ}$~. Baryon number is conserved
and, thus, proton is stable as automatic consequences of 
a R-symmetry. Light neutrinos are 
assumed to acquire degenerate masses equal to about 
$1.5~{\rm{eV}}$ through their couplings to $SU(2)_L$ 
triplet superfields. These neutrinos constitute the hot 
dark matter of the universe. After inflation, the inflaton 
system decays into $SU(2)_L$ triplets which subsequently  
decay producing a primordial lepton asymmetry later 
converted into the observed baryon asymmetry of the universe 
by sphaleron effects. The gravitino and baryogenesis 
constraints can be satisfied with `natural' values 
($\sim 10^{-3}$) of the relevant coupling constants.

\vspace{0.5cm}
This work was supported by the research grant 
PENED/95 K.A.1795.

\def\ijmp#1#2#3{{ Int. Jour. Mod. Phys. }{\bf #1~}(19#2)~#3}
\def\pl#1#2#3{{ Phys. Lett. }{\bf B#1~}(19#2)~#3}
\def\zp#1#2#3{{ Z. Phys. }{\bf C#1~}(19#2)~#3}
\def\prl#1#2#3{{ Phys. Rev. Lett. }{\bf #1~}(19#2)~#3}
\def\rmp#1#2#3{{ Rev. Mod. Phys. }{\bf #1~}(19#2)~#3}
\def\prep#1#2#3{{ Phys. Rep. }{\bf #1~}(19#2)~#3}
\def\pr#1#2#3{{ Phys. Rev. }{\bf D#1~}(19#2)~#3}
\def\np#1#2#3{{ Nucl. Phys. }{\bf B#1~}(19#2)~#3}
\def\mpl#1#2#3{{ Mod. Phys. Lett. }{\bf #1~}(19#2)~#3}
\def\arnps#1#2#3{{ Annu. Rev. Nucl. Part. Sci. }{\bf
#1~}(19#2)~#3}
\def\sjnp#1#2#3{{ Sov. J. Nucl. Phys. }{\bf #1~}(19#2)~#3}
\def\jetp#1#2#3{{ JETP Lett. }{\bf #1~}(19#2)~#3}
\def\app#1#2#3{{ Acta Phys. Polon. }{\bf #1~}(19#2)~#3}
\def\rnc#1#2#3{{ Riv. Nuovo Cim. }{\bf #1~}(19#2)~#3}
\def\ap#1#2#3{{ Ann. Phys. }{\bf #1~}(19#2)~#3}
\def\ptp#1#2#3{{ Prog. Theor. Phys. }{\bf #1~}(19#2)~#3}
\def\plb#1#2#3{{ Phys. Lett. }{\bf#1B~}(19#2)~#3}
\def\apjl#1#2#3{{Astrophys. J. Lett. }{\bf #1~}(19#2)~#3}
\def\n#1#2#3{{ Nature }{\bf #1~}(19#2)~#3}
\def\apj#1#2#3{{Astrophys. Journal }{\bf #1~}(19#2)~#3}
\def\ibid#1#2#3{{ibid. }{\bf #1~}(19#2)~#3}


\begin{references}

\bibitem{structure} Q. Shafi and F. W. Stecker, 
\prl{53}{84}{1292}. For a recent review and other references 
see Q. Shafi and R. K. Schaefer, hep-ph/9612478.

\bibitem{lyth} E. J. Copeland, A. R. Liddle, D. H. Lyth, 
E. D. Stewart and D. Wands, \pr{49}{94}{6410}.

\bibitem{dss} G. Dvali, Q. Shafi and R. Schaefer, 
\prl{73}{94}{1886}.
 
\bibitem{hybrid}  A. D. Linde, \pl{259}{91}{38}; 
\pr{49}{94}{748}.

\bibitem{dls} G. Dvali, G. Lazarides and Q. Shafi, 
\pl{424}{98}{259}.

\bibitem{gravitino} M. Yu. Khlopov and A. D. Linde, 
\plb{138}{84}{265}; J. Ellis, J. E. Kim and D. Nanopoulos, 
\plb{145}{84}{181}.

\bibitem{atmospheric} G. Lazarides and N. D. Vlachos, 
\pl{441}{98}{46}.

\bibitem{kn} J. E. Kim and H. P. Nilles, 
\plb{138}{84}{150}. 

\bibitem{rsym} G. Lazarides and Q. Shafi, 
\pr{58}{98}{071702}.

\bibitem{pq} R. Peccei and H. Quinn, \prl{38}{77}{1440}; 
S. Weinberg, \prl{40}{78}{223}; F. Wilczek, 
\prl{40}{78}{279}.

\bibitem{superk} T. Kajiata, talk given at the XVIIIth 
International Conference on Neutrino Physics and 
Astrophysics (Neutrino'98), Takayama, Japan, 4-9 June, 1998.

\bibitem{triplet} G. Lazarides, Q. Shafi and C. Wetterich, 
\np{181}{81}{287}; C. Wetterich, \np{187}{81}{343}; 
R. N. Mohapatra and G. Senjanovic, \pr{23}{81}{165}; 
R. Holman, G. Lazarides and Q. Shafi, \pr{27}{83}{995}.

\bibitem{sarkar} U. Sarkar, hep-ph/9807466.

\bibitem{leptogenesis} M. Fukugita and T. Yanagida, 
\pl{174}{86}{45}; W. Buchm\"uller and M. Pl\"umacher, 
\pl{389}{96}{73}. In the context of inflation see 
G. Lazarides and Q. Shafi, \pl{258}{91}{305}; 
G. Lazarides, C. Panagiotakopoulos and Q. Shafi, 
\pl{315}{93}{325}, (E) \ibid{317}{93}{661}. 

\bibitem{lss} G. Lazarides, R. Schaefer and Q. Shafi, 
\pr{56}{97}{1324}.

\bibitem{cobe} G. F. Smoot {\it et al.}, \apjl{396}
{92}{L1}; C. L. Bennett {\it et al.}, \apjl{464}{96}{1}.

\bibitem{gg} H. Georgi and S. L. Glashow, hep-ph/9808293.

\bibitem{bimaximal} V. Barger, S. Pasvaka, T. Weiler 
and K. Whisnant, hep-ph/9806387.

\bibitem{vacuum} V. Barger, R. J. N. Phillips and 
K. Whisnant, \pr{24}{81}{538}; S. L. Glashow and 
L. M. Krauss, \pl{190}{87}{199}.

\bibitem{beta} L. Baudis {\it{et al.}}, \pl{407}{97}{219}.

\bibitem{mohapatra} R. N. Mohapatra and S. Nussinov,
hep-ph/9809415.

\bibitem{photon} M. Kawasaki and T. Moroi, \ptp{93}
{95}{879}. 

\bibitem{covi} L. Covi, E. Roulet and F. Vissani, 
\pl{384}{96}{169}. 

\bibitem{pilaftsis} A. Pilaftsis, \pr{56}{97}{5431}.

\bibitem{ibanez} L. E. Ib\'a\~nez and F. Quevedo, 
\pl{283}{92}{261}.

\bibitem{deuterium} D. Tytler, X. M. Fan and S. Burles, 
\n{381}{96}{207}; S. Burles and D. Tytler, \apj{460}{96}
{584}.

\end{references}
\end{document}